\documentclass[aps, prd, amsmath, floats, floatfix, twocolumn, nofootinbib, superscriptaddress, showpacs]{revtex4-1}
\usepackage{graphicx}
\usepackage{color}
\usepackage{latexsym}
\usepackage{amsfonts}

\newcommand{\be}{\begin{eqnarray}}
\newcommand{\ee}{\end{eqnarray}}

\begin{document}

\title{Iron K$\alpha$ line of Proca stars}

\author{Tianling Shen}
\affiliation{Center for Field Theory and Particle Physics and Department of Physics, Fudan University, 200433 Shanghai, China}

\author{Menglei Zhou}
\affiliation{Center for Field Theory and Particle Physics and Department of Physics, Fudan University, 200433 Shanghai, China}

\author{Cosimo Bambi}
\email[Corresponding author: ]{bambi@fudan.edu.cn}
\affiliation{Center for Field Theory and Particle Physics and Department of Physics, Fudan University, 200433 Shanghai, China}
\affiliation{Theoretical Astrophysics, Eberhard-Karls Universit\"at T\"ubingen, 72076 T\"ubingen, Germany}

\author{Carlos A. R. Herdeiro}
\affiliation{Departamento de F\`isica da Universidade de Aveiro and Center for Research and Development in Mathematics and Applications (CIDMA), Campus de Santiago, 3810-183 Aveiro, Portugal}

\author{Eugen Radu}
\affiliation{Departamento de F\`isica da Universidade de Aveiro and Center for Research and Development in Mathematics and Applications (CIDMA), Campus de Santiago, 3810-183 Aveiro, Portugal}

\date{January 2017}

\begin{abstract}
X-ray reflection spectroscopy can be a powerful tool to test the nature of astrophysical black holes. Extending previous work on Kerr black holes with scalar hair~\cite{p2} and on boson stars~\cite{p3}, here we study whether astrophysical black hole candidates may be horizonless, self-gravitating, vector Bose-Einstein condensates, known as \textit{Proca stars}~\cite{proca}. We find that observations with current X-ray missions can only provide weak constraints and rule out solely Proca stars with low compactness. There are two reasons. First, at the moment we do not know the geometry of the corona, and therefore the uncertainty in the emissivity profile limits the ability to constrain the background metric. Second, the photon number count is low even in the case of a bright black hole binary, and we cannot have a precise measurement of the spectrum. 
\end{abstract}

\maketitle


\section{Introduction}

Today we have robust observational evidence for the existence of dark and compact objects that can be naturally interpreted as black holes (BHs)  -- see $e.g.$~\cite{Narayan:2013gca}. These BH candidates could, however, be something else, but only in the presence of new physics~\cite{rmp}. Stellar-mass BH candidates have a mass $M > 3$~$M_\odot$, and are therefore too heavy to be compact relativistic stars, at least if composed by standard matter~\cite{rr}. Supermassive BH candidates of $10^5$-$10^{10}$~$M_\odot$ are at the center of galaxies and are too massive, compact, and old to be clusters of non-luminous bodies~\cite{maoz}. Moroever, the gravitational waves observed by LIGO are consistent with the signal expected from the coalescence of two BHs in general relativity~\cite{yyp}, even though we have not yet entered the era of precision gravitational wave spectroscopy.

\bigskip

Within the framework of standard physics, the spacetime metric around astrophysical BHs should be well approximated by the Kerr solution. Nevertheless, macroscopic deviations are predicted in a number of scenarios involving new physics, which typically can be divided into two classes: $(i)$ modified gravity; or $(ii)$ exotic matter (within general relativity). In the latter context, one of the most natural and interesting models for BH mimickers is that of a horizonless, self-gravitating Bose-Einstein condensate of ultra-light bosons. The first such model was found long ago by Kaup~\cite{Kaup:1968zz} and Ruffini and Bonazzola~\cite{Ruffini:1969qy}, corresponding to \textit{scalar boson stars}. The model requires only a minimal number of (physically reasonable) new ingredients:  a massive complex scalar field (no self-interactions are required, even though they are possible~\cite{Colpi:1986ye,Herdeiro:2015tia}) minimally coupled to Einstein's gravity; moroever, an open set of the domain of existence of these boson stars are known to be stable and even form dynamically (see~\cite{Schunck:2003kk,Liebling:2012fv} for reviews). More recently, spin 1 cousins of these scalar boson stars, named \textit{Proca stars}, have also been found in Einstein's gravity minimally coupled to a massive complex vector field~\cite{proca}. As for their scalar cousins, these solution can be either static or rotating and seem to mimic some, but not all, properties of the scalar case~\cite{Herdeiro:2016tmi}.

\bigskip

In this letter, we consider the Proca star solutions found in Ref.~\cite{proca}. In particular, we want to address the questions whether Proca stars can be a viable alternative to BHs to explain the observed dark and compact objects in the Universe or, on the contrary, can be already ruled out by current astrophysical observations. We answer this question by considering the shape of the iron K$\alpha$ line commonly observed in the reflection spectra of astrophysical BH candidates. We find that non-compact configurations are not consistent with the available X-ray data, but the constraints are weak and more compact Proca stars can well mimic Kerr BHs.

\section{Proca stars}
Vector boson stars (a.k.a. Proca stars) are solutions to Einstein's gravity with a minimally coupled complex Proca field of mass $\mu$~\cite{proca}. The action of the model reads
\be
\mathcal{S} = \mathcal{S}_{\rm EH} + \int d^4x \sqrt{-g} \, \mathcal{L}_{\rm Proca} \, ,
\ee
where $\mathcal{S}_{\rm EH}$ is the Einstein-Hilbert action and
\be
\mathcal{L}_{\rm Proca} = 
- \frac{1}{4} F_{\alpha\beta} \bar{F}^{\alpha\beta} 
- \frac{1}{2} \mu^2 A_\alpha \bar{A}^\alpha \, ,
\ee
where $A$ is the (complex) potential 1-form, $F = dA$ is the field strength, and the overbar denotes the complex conjugate quantities.

With appropriate anstaz~\cite{proca}, both static, spherically symmetric and stationary, rotating solution can be found, corresponding to macroscopic, self-gravitating lumps of the Proca field. These are prevented from gravitationally collapsing by an effective pressure associated to a harmonic time dependence in the potential 1-form, $A\sim e^{-iwt+im\varphi}$, where $t$ and $\phi$ are the coordinates associated to the timelike (at infinity) Killing vector field $\partial/\partial t$ and azimuthal Killing vector field $\partial/\partial \phi$. $w \in \mathbb{R}^+$ is the frequency of harmonic oscillation and $m\in \mathbb{Z}/ \{0\}$ is the azimuthal quantum number. Observe that due to the complex nature of the field, both the $t$ and $\phi$ dependence of the Proca potential ansatz vanish at the level of the energy-momentum tensor, making this ansatz compatible with a stationary and axi-symmetric geometry~\cite{Herdeiro:2015gia}.

\begin{figure}[t]
\begin{center}
\includegraphics[type=pdf,ext=.pdf,read=.pdf,width=8.0cm]{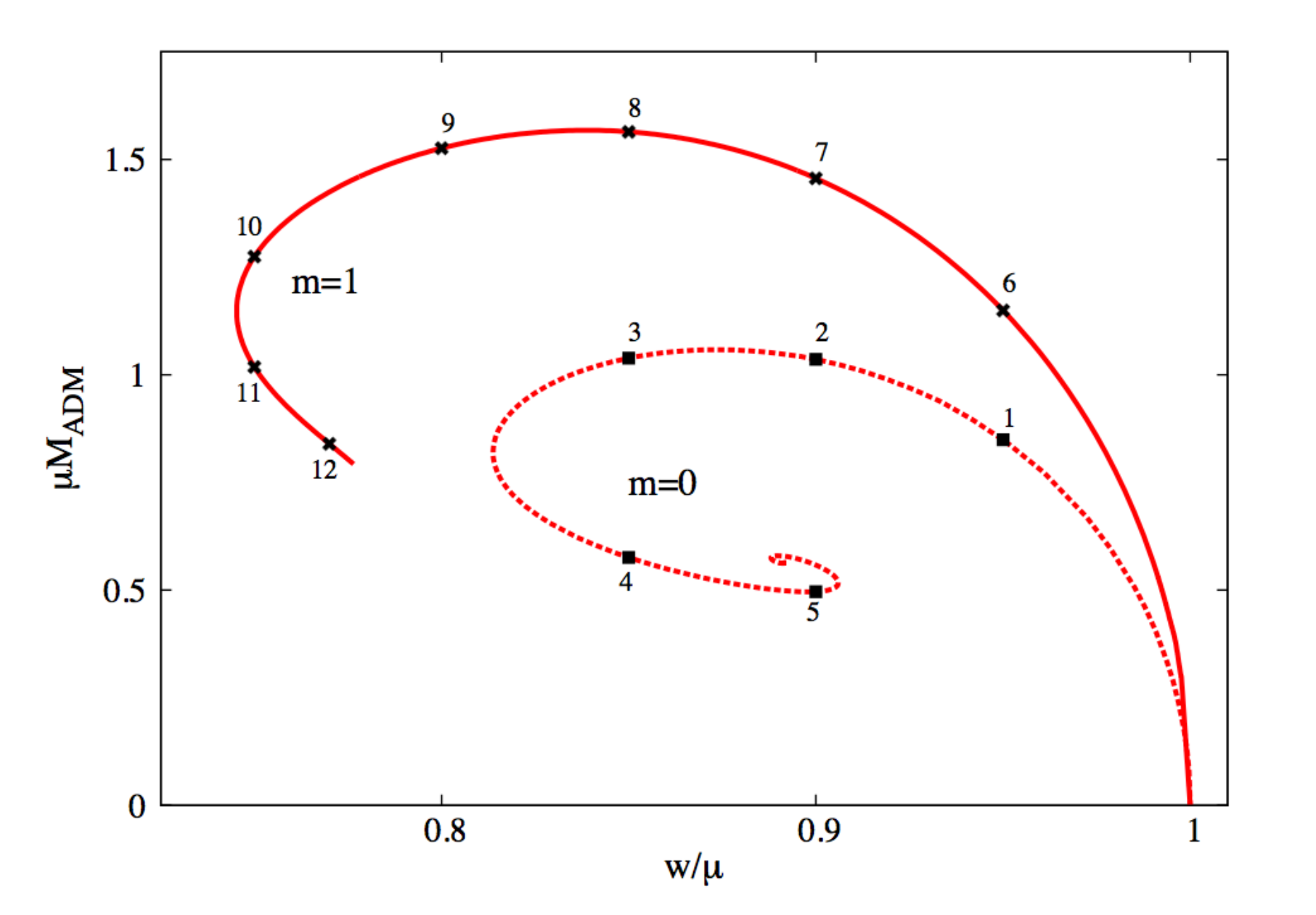}
\end{center}
\vspace{-0.5cm}
\caption{Proca star solutions in an ADM mass $vs.$ vector field frequency diagram. The red dashed line describes the family of spherical Proca stars with $m=0$, while the red solid line is for the family of rotating Proca stars with $m=1$. The 12~highlighted points correspond to the configurations studied in this work. \label{f-proca}}
\end{figure}

In Fig.~\ref{f-proca} we exhbit the existence domain for spherically symmetric ($m=0$) and rotating (with $m=1$) Proca stars, in an ADM mass $vs.$ Proca field frequency diagram, both quantities being made dimensionless by using the Proca field mass $\mu$. Both types of solutions fall along a spiral-type line in this diagram, similarly to what happens to scalar boson stars and also for solutions with higher $m$~\cite{proca}. As a rule of thumb, the compactness of the solutions typically increases as we move along this spiral, starting from the vacuum case (maximal value of $w$, minimal value of $M$) -- see Fig. 2 in~\cite{Herdeiro:2015gia} for a more detailed analysis of compactness, for the case of scalar boson stars. Also, the stability of the solutions depends on their location along the spiral. For the spherically symmetric case it was shown in~\cite{proca} that the set of solutions starting at the vacuum point and up to the maximal mass are stable; at the maximal ADM mass an unstable mode develps and solutions beyond this point are expected to be perturbatively unstable. For the rotating case, albeit no detailed perturbative computation has been made, some generic arguments suggest a similar picture should hold (see Sec. 6.2 in~\cite{Herdeiro:2015gia} for such a discussion in the case of scalar boson stars). 

In Fig.~\ref{f-proca}  we have highlighted 12 representative solutions, 5 for the spherically symmetric case and 7 for the rotating case, corresponding to the sample of Proca stars that will be studied in detail in this paper. Some properties of these solutions are detailed in Table I. As a relevant property for the following we observe that timelike, co-rotating, circular, equatorial geodesics exist all the way up to the origin for all solutions studied. In other words there is no innermost (stable or unstable) circular geodesic, for co-rotating orbits (the counter-rotating case is more complicated and will be detailed elsewhere).

We remark that Proca (scalar boson) stars can be continuously connected to Kerr BHs via a larger family of solutions: Kerr BHs with Proca~\cite{Herdeiro:2016tmi} (scalar~\cite{Herdeiro:2014goa}) hair, showing there is a general pattern and a general mechanism at work~\cite{Herdeiro:2014ima}.

\begin{table}[t]
\begin{tabular}{|c|cccc|}
\hline
\hspace{0.2cm} Solution \hspace{0.2cm} & \hspace{0.2cm} $w$ \hspace{0.2cm} & \hspace{0.1cm} branch \hspace{0.1cm} & \hspace{0.3cm} $M$ \hspace{0.3cm} & \hspace{0.3cm} $J$ \hspace{0.3cm} \\
\hline 
1 & 0.95 & 1st & 0.849 & 0 \\
2 & 0.90 & 1st & 1.036 & 0 \\
3 & 0.85 & 1st & 1.039 & 0 \\
4 & 0.85 & 2nd & 0.576 & 0 \\
5 & 0.90 & 2nd & 0.496 & 0 \\
\hline
6 & 0.95 & 1st & 1.149 & 1.171 \\
7 & 0.90 & 1st & 1.456 & 1.500 \\
8 & 0.85 & 1st & 1.564 & 1.622 \\
9 & 0.80 & 1st & 1.526 & 1.574 \\
10 & 0.75 & 1st & 1.274 & 1.247 \\
11 & 0.75 & 2nd & 1.018 & 0.904 \\
12 & 0.77 & 2nd & 0.840 & 0.668 \\
\hline
\end{tabular}
\vspace{0.4cm}
\caption{Properties of the Proca star solutions 1-12. \label{tab}}
\end{table}

\section{Reflection spectrum}

Broad iron lines are a common feature in the X-ray spectrum of both stellar-mass and supermassive BH candidates. In the disk-corona model~\cite{corona1,corona2}, a BH is surrounded by a geometrically thin and optically thick accretion disk. The disk emits like a blackbody locally and a multi-color blackbody when integrated radially. The corona is a hot ($\sim 100$~keV), usually optically thin, electron cloud. For instance, it may be the base of the jet or a cloud covering the BH or the inner part of the accretion disk, but the actual geometry is currently unknown. Due to inverse Compton scattering of thermal photons from the disk off hot electrons in the corona, the latter becomes an X-ray source with a power-law spectrum $E^{-\Gamma}$, where $\Gamma \approx 1$-3. The corona illuminates the disk, producing a reflection component with some fluorescent emission lines. The most prominent feature of the reflection spectrum is usually the iron K$\alpha$ line, which is at about 6.4~keV in the case of neutral iron atoms and shifts up to 6.97~keV in the case of H-like iron ions.

Iron K$\alpha$ lines in the reflection spectrum of BHs are broad and skewed as a result of relativistic effects (Doppler boosting, gravitational redshift, light bending) occurring in the strong gravitational field of the source. Assuming the Kerr metric, the analysis of the iron K$\alpha$ line can be used to measure the BH spin parameter~\cite{iron1,iron2}. Relaxing the Kerr BH hypothesis, the technique can probe the metric around the compact object~\cite{iron3,iron4}. Actually one has to analyze the whole reflection spectrum, not only the iron line, but most of the information about the spacetime metric in the strong gravity region is in the iron line and for this reason the technique is often referred to as the iron line method. It is remarkable that, in the presence of high quality data and the correct astrophysical model, this approach can be a powerful tool to test the nature of BHs~\cite{icb1,icb2,icb3}.

The shape of the iron K$\alpha$ line is determined by the metric of the spacetime, the inclination angle of the disk with respect to the line of sight of the distant observer $i$, the geometry of the emitting region, and the emissivity profile of the disk. The emission is usually assumed from the inner edge of the disk $r_{\rm in}$ to some large outer radius $r_{\rm out}$, but the exact value of the latter is not important because the emissivity is lower and lower as the radius increases. One usually tries to select the sources in which the inner edge of the disk is at the radius of the innermost stable circular orbit (ISCO). In the case of a corona with arbitrary geometry, it is common to model the emissivity profile with a broken power-law, namely to assume that the emissivity scales as $1/r^{q_1}$ for $r <  r_{\rm b}$ and as $1/r^{q_2}$ for $r >  r_{\rm b}$, where the emissivity indices $q_1$ and $q_2$ and the breaking radius $r_{\rm b}$ are three free parameters to be determined by the fit.

In the case of the spacetimes of Proca stars, there is no ISCO for co-rotating orbits, so the possible accretion disk may either extend up to the center of the object or be truncated at some small radius. In our simulations, we assume the former scenario and we employ the following lamppost-inspired emissivity profile~\cite{dauser}:
\be\label{emissivity}
I \propto \frac{h}{(r^2 + h^2)^{3/2}} \, , 
\ee
where $h$ is the height of the corona along the spin axis of the BH in the lamppost set-up and in our case we choose $h=2$ (in units in which $1/\mu = 1$). The shapes of the expected iron lines in the reflection spectrum of the Proca stars with $m=0$ (solutions 1-5) and $m=1$ (solutions 6-12) are shown, respectively, in the left and right panels in Fig.~\ref{f-line} for $i=45^\circ$. The calculations are done with the code described in Refs.~\cite{c1,c2}.

\section{Simulations}

We want now to address the question whether current observations of the reflection spectrum of astrophysical BH candidates can rule out, or constrain, the possibility that these objects are actually Proca stars. As an explorative study, we do not consider specific observations. We instead follow the strategy already employed in Refs.~\cite{p1,p2,p3,p4}, which permits to get quickly a rough estimate of current constraints. The key point is that current observations are consistent with the Kerr metric, in the sense that X-ray data are normally fitted with reflection spectra computed in the Kerr metric and the result is acceptable. We can thus simulate some observations of the X-ray spectrum of Proca stars and then try to fit the simulated data with a model calculated in the Kerr background. If the fit is acceptable, we can say that current data cannot rule out the Proca star solution of that simulation. If the fit is bad, we can say that our Proca star solution cannot describe the metric around the observed astrophysical BH candidates, because there is currently no tension between observations and theoretical models.

We simulate observations with XIS/Suzaku\footnote{http://heasarc.gsfc.nasa.gov/docs/suzaku/} of a bright BH binary. For simplicity, we model the spectrum of the source with a power-law with photon index $\Gamma = 2$ (representing the primary spectrum of the corona) and a single iron line (the reflection spectrum of the disk). We assume typical parameters for a bright BH binary. The energy flux in the 0.7-10~keV range is about $4 \cdot 10^9$~erg/s/cm$^2$. The equivalent width of the iron line is about 200~eV. We assume that the exposure time is 100~ks and the photon count turns out to be about $3 \cdot 10^7$. For every Proca star solutions, we consider three viewing angles, namely $i = 20^\circ$, $45^\circ$, and $70^\circ$, but we find that the final result is not very sensitive to $i$.

We treat these simulations as real data. We use XSPEC\footnote{http://heasarc.gsfc.nasa.gov/docs/xanadu/xspec/index.html} and fit the simulated data with a power-law and a Kerr iron line. The latter is modeled with RELLINE~\cite{relline}. We have eight free parameters in the fit: the photon index of the power-law $\Gamma$, the normalization of the power-law component, the spin of the BH $a_*$, the viewing angle $i$, the two indices $q_1$ and $q_2$, the breaking radius $r_{\rm b}$, and the normalization of the iron line.

\begin{figure*}[t]
\begin{center}
\includegraphics[type=pdf,ext=.pdf,read=.pdf,width=8.9cm]{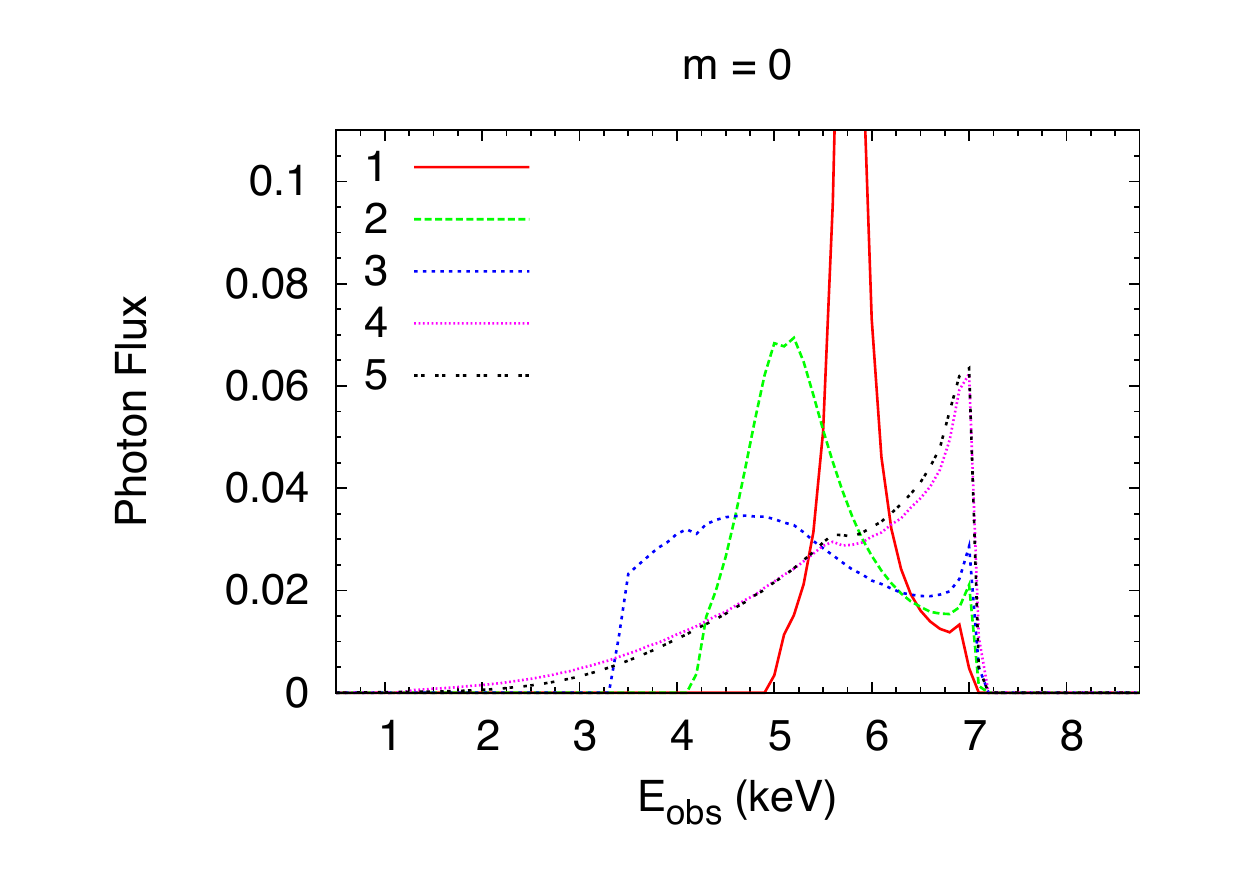}
\hspace{-0.5cm}
\includegraphics[type=pdf,ext=.pdf,read=.pdf,width=8.9cm]{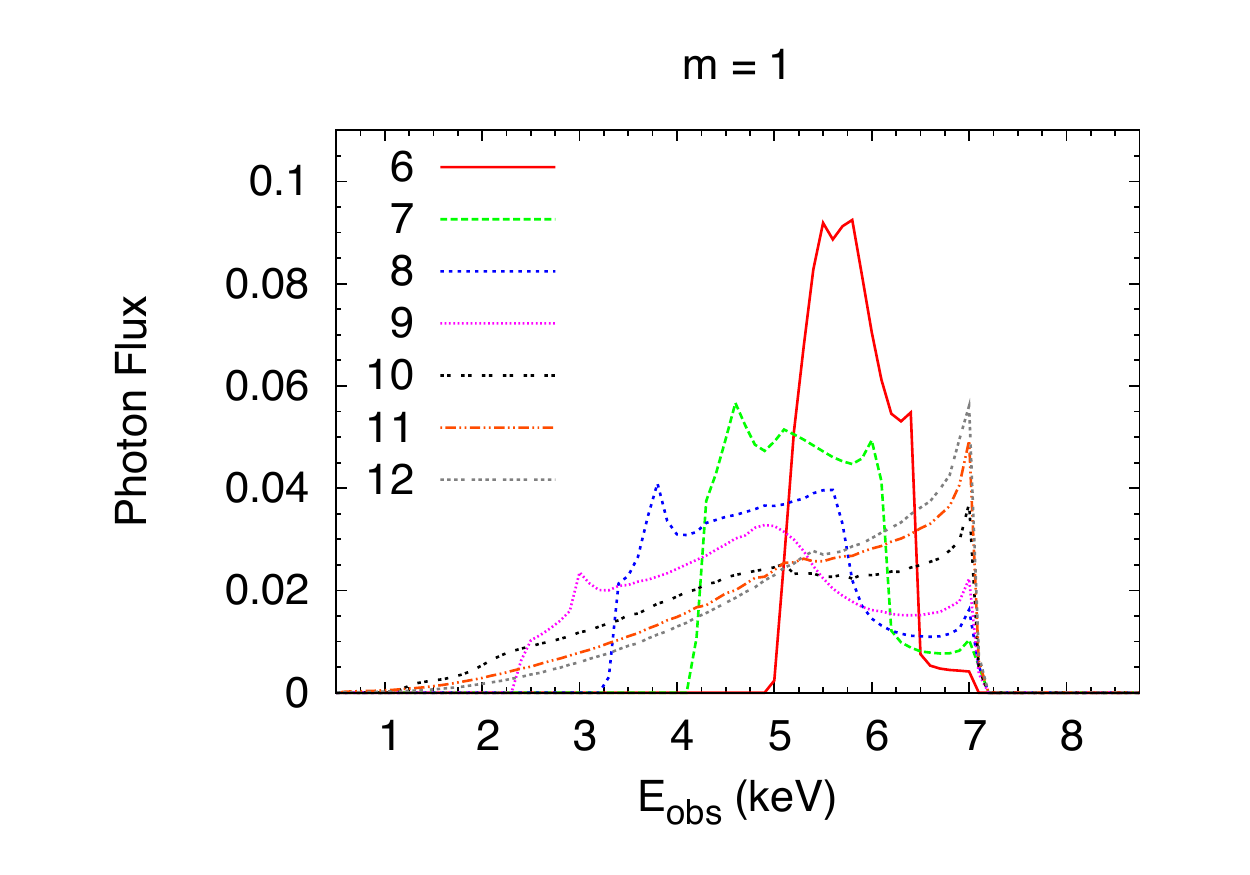}
\end{center}
\vspace{-0.5cm}
\caption{Expected shape of single iron lines in the reflection spectrum of Proca stars with $m=0$ (solutions 1-5, left panel) and $m=1$ (solutions 6-12, right panel). The viewing angle is $i = 45^\circ$ and the emissivity profile is $\propto h/(r^2 + h^2)^{3/2}$ with $h = 2$. See the text for more details. \label{f-line}}
\vspace{0.5cm}
\begin{center}
\includegraphics[type=pdf,ext=.pdf,read=.pdf,width=8.0cm]{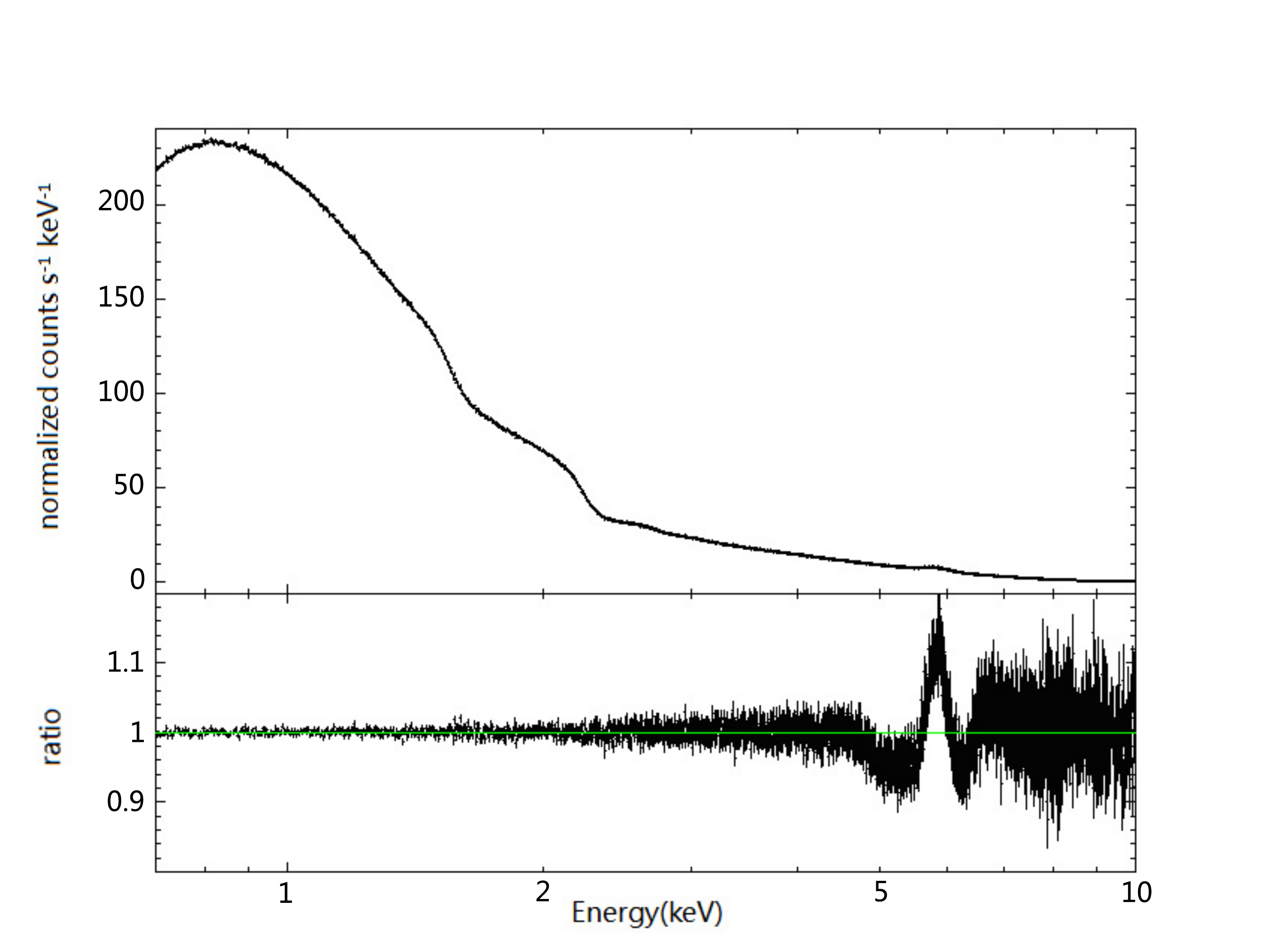}
\includegraphics[type=pdf,ext=.pdf,read=.pdf,width=8.0cm]{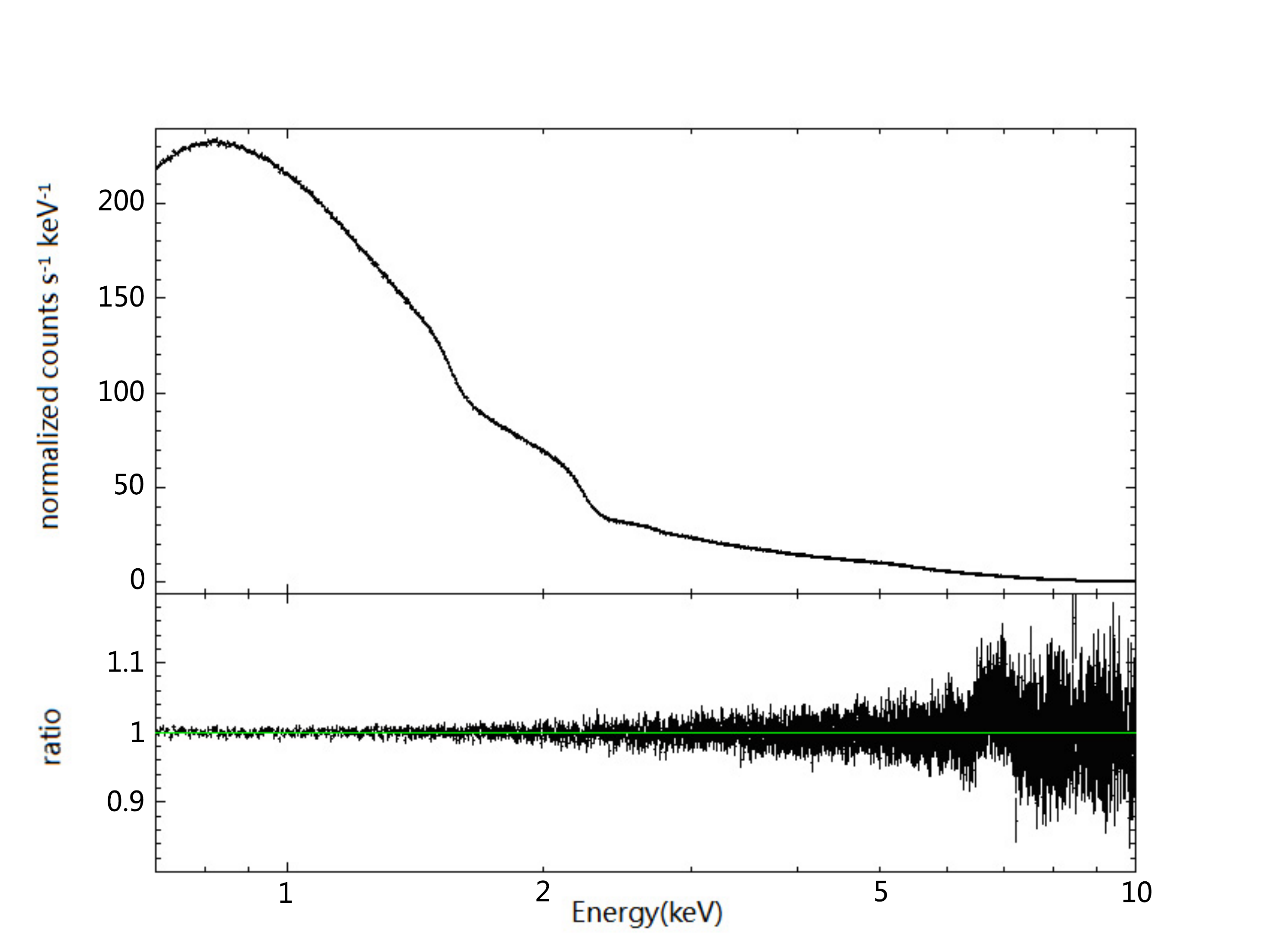} \\
\includegraphics[type=pdf,ext=.pdf,read=.pdf,width=8.0cm]{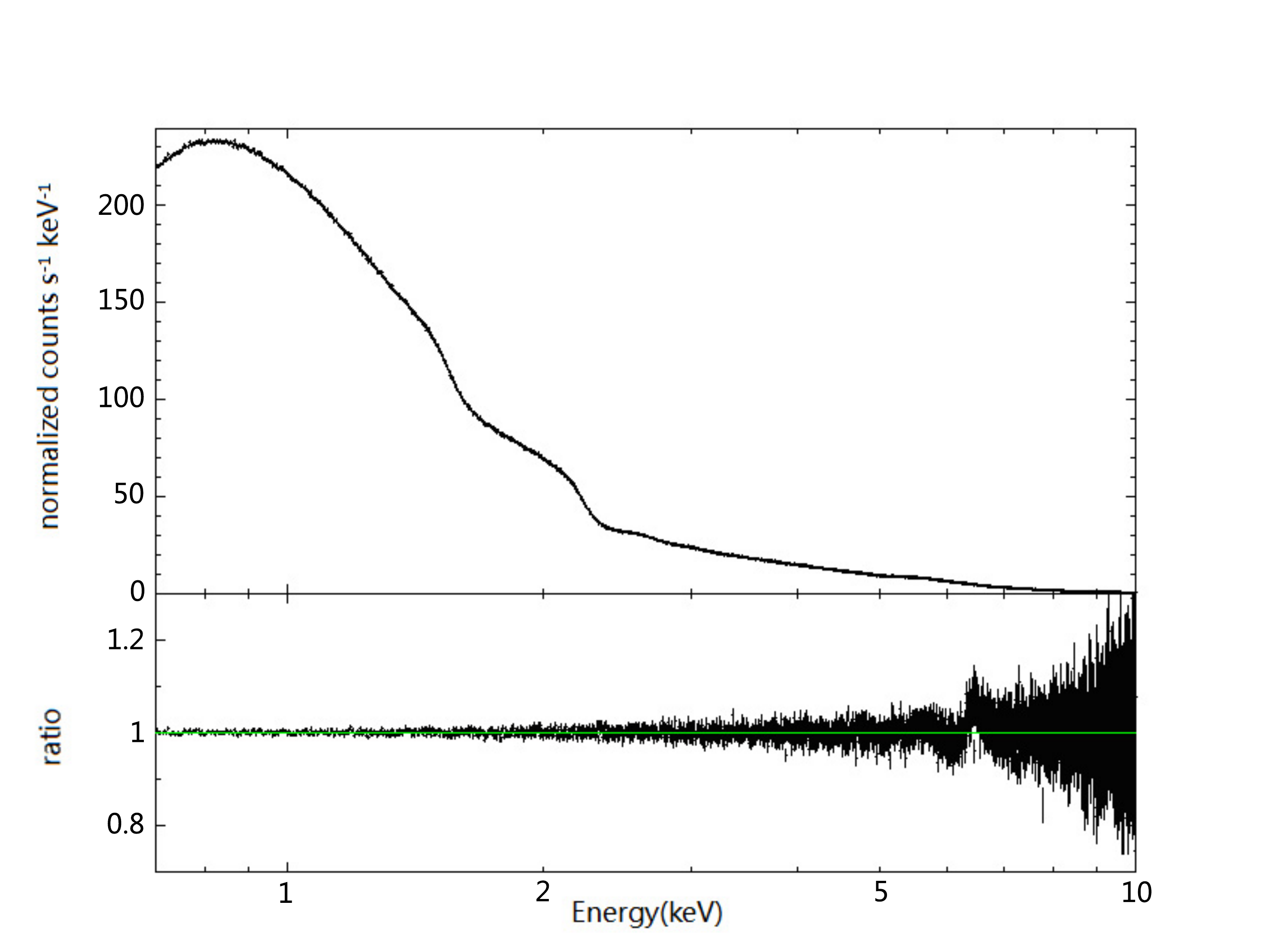}
\includegraphics[type=pdf,ext=.pdf,read=.pdf,width=8.0cm]{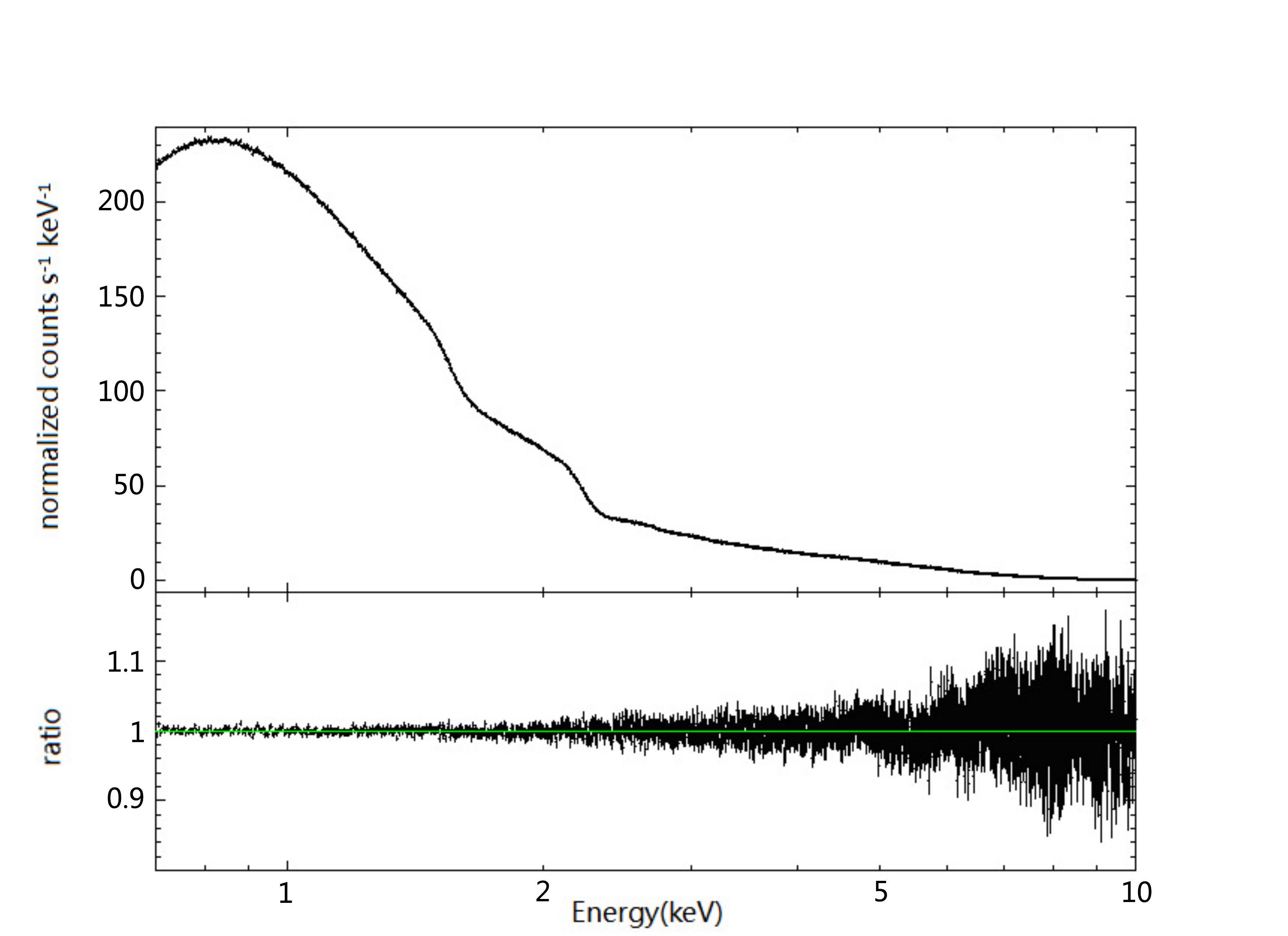}
\end{center}
\caption{Results of our simulations and fits for Proca stars solution~1 (top left panel), solution~2 (top right panel), solution~6 (bottom left panel), and solution~7 (bottom right panel). In every panel, the top plot shows the simulated data and the best fit, while the bottom panel shows their ratio. See the text for more details. \label{f-sim}}
\end{figure*}

The results of our simulations for the 12 Proca star solutions can be summarized as follows. For $m=0$, solution~1 cannot be fitted with a Kerr model, solution~2 is marginally consistent, while solutions~3-5 can be well fitted with a Kerr model (even if we obtain completely wrong estimates of some parameters). For $m=1$, we have a similar situation. Solution~6 cannot be well fitted with a Kerr model, for solution~7 the fit is already acceptable, solutions~8-12 can be well fitted with a Kerr model. Fig.~\ref{f-sim} shows the results for solutions~1, 2, 6, and 7 in the case $i = 45^\circ$. In every panel, the top plot shows the simulated data and the best fit (folded spectra). The bottom panel shows the ratio between the simulated data and the best fit, which is the key-point in our simple analysis. If the ratio is close to 1, in every region of the spectrum and within the error bars of the measurement, the model can well fit the data. This is clearly not the case for solution~1 (top left panel) and solution~6 (bottom left panel) around 6~keV. In the case of solution~2 (top right panel) and solution~7 (bottom right panel), we see that at some energies most of the data are above or below the line of ratio equal 1, but the error bars are large enough that the ratio 1 is included, and for solution~7 the fit is surely acceptable. For a longer exposure time, the size of the error bars would decrease and show clearly if the Kerr model cannot fit the data.

As explained in the previous section, in our simulations the accretion disk extends untill the center at $r=0$, because there is no ISCO in these spacetimes for co-rotating particles. If we assume that the disk is truncated at some very small radius, the resulting iron line is not very different within the emissivity profile in Eq.~(\ref{emissivity}) and our conclusions are unchanged. If we move the inner edge of the disk to larger radii, we cannot have the low energy tail in the iron line profile and it is easier to rule out these spacetimes. Our conclusions are thus based on the simplest choice, which provides the most conservative bounds.

\section{Concluding remarks}

In this letter, we have extended previous work on Kerr BHs with scalar hair~\cite{p2} and boson stars~\cite{p3} to the case of the Proca star solutions found in Ref.~\cite{proca}. We have studied if current X-ray observations of the reflection spectrum of stellar-mass BH candidates may be consistent with that expected for a Proca star or, otherwise, current observations can already rule out Proca stars as alternative to BHs.

Our simulations show that current constraints on the existence domain of Proca stars are weak. Still, we can rule out non-compact objects (solution~1 for $m=0$ and solution~6 for $m=1$, corresponding also to the largest frequencies $w$), while more compact configurations can mimic Kerr BHs and have reflection spectra with broad iron lines. But observe that some of the most compact configurations are (or are \textit{likely}, in the rotating case) unstable, and hence may be ruled out on different grounds.

Our constraints are weak for two reasons. First, the uncertainty in the emissivity profile limits the ability to perform precise measurements of the background metric. Second, with current X-ray missions the statistics in the iron line is not good enough. These two points are not only true in the case of Proca stars, but for tests of the Kerr metric using X-ray reflection spectroscopy in general, and the results of this work confirm that at the moment these are the two weak points of this approach. Much stronger constraints should be possible with the next generation of X-ray missions. Timing measurements will be able to test the exact geometry of the corona, which may permit to have a theoretical prediction of the emissivity profile. The much larger effective area of the next generation of X-ray missions will permit to have a sufficiently high photon number count to get precise measurements of the spectra of the sources.

To conclude, the analysis in this and our previous papers~\cite{p2,p3}, establishes that with present and forthcoming astrophysical observations,  the iron line technique can be used to set informative constraints on the domain of existence of boson stars and their hairy BH counterparts. Together with other astrophysical information, such as that obtained from gravitational lensing~\cite{Cunha:2015yba,Vincent:2016sjq} and quasi-periodic oscillations~\cite{Franchini:2016yvq}, these fairly simple alternatives to the Kerr BH paradigm, which occur within general relativity albeit involving non-standard model matter, can undergo precision testing in the forthcoming years.


\begin{acknowledgments}
The work of T.S., M.Z., and C.B. was supported by the NSFC (grants U1531117 and 11305038) and the Thousand Young Talents Program. C.B. also acknowledges the support from the Alexander von Humboldt Foundation. C.A.R.H. and E.R. acknowledge funding from the FCT-IF programme. This work was partially supported by the H2020-MSCA-RISE-2015 Grant No. StronGrHEP-690904, and by the CIDMA project UID/MAT/04106/2013.
\end{acknowledgments}



\begin{thebibliography}{99}

\bibitem{p2} 
  Y.~Ni, M.~Zhou, A.~Cardenas-Avendano, C.~Bambi, C.~A.~R.~Herdeiro and E.~Radu,
  JCAP {\bf 1607}, 049 (2016)
  [arXiv:1606.04654 [gr-qc]].  
  
\bibitem{p3} 
  Z.~Cao, A.~Cardenas-Avendano, M.~Zhou, C.~Bambi, C.~A.~R.~Herdeiro and E.~Radu,
  JCAP {\bf 1610}, 003 (2016)
  [arXiv:1609.00901 [gr-qc]].  
  
  \bibitem{proca} 
  R.~Brito, V.~Cardoso, C.~A.~R.~Herdeiro and E.~Radu,
  Phys.\ Lett.\ B {\bf 752}, 291 (2016)
  [arXiv:1508.05395 [gr-qc]].  
  
\bibitem{Narayan:2013gca} 
  R.~Narayan and J.~E.~McClintock,
  arXiv:1312.6698 [astro-ph.HE].

\bibitem{rmp}
  C.~Bambi,
  Rev.\ Mod.\ Phys.\  (in press)
  [arXiv:1509.03884 [gr-qc]].

\bibitem{rr}
  C.~E.~Rhoades and R.~Ruffini,
  Phys.\ Rev.\ Lett.\  {\bf 32}, 324 (1974). 
   
\bibitem{maoz} 
  E.~Maoz,
  Astrophys.\ J.\  {\bf 494}, L181 (1998)
  [astro-ph/9710309].

\bibitem{yyp} 
  N.~Yunes, K.~Yagi and F.~Pretorius,
  Phys.\ Rev.\ D {\bf 94}, 084002 (2016)
  [arXiv:1603.08955 [gr-qc]].

\bibitem{Kaup:1968zz} 
  D.~J.~Kaup,
  Phys.\ Rev.\  {\bf 172}, 1331 (1968).

\bibitem{Ruffini:1969qy}
  R.~Ruffini and S.~Bonazzola,
  Phys.\ Rev.\  {\bf 187}, 1767 (1969).
  
\bibitem{Colpi:1986ye} 
  M.~Colpi, S.~L.~Shapiro and I.~Wasserman,
  Phys.\ Rev.\ Lett.\  {\bf 57}, 2485 (1986).

\bibitem{Herdeiro:2015tia} 
  C.~A.~R.~Herdeiro, E.~Radu and H.~Rœnarsson,
  Phys.\ Rev.\ D {\bf 92}, no. 8, 084059 (2015)
  [arXiv:1509.02923 [gr-qc]].

  
\bibitem{Schunck:2003kk}
  F.~E.~Schunck and E.~W.~Mielke,
  Class.\ Quant.\ Grav.\  {\bf 20}, R301 (2003)
  [arXiv:0801.0307 [astro-ph]].

\bibitem{Liebling:2012fv}
  S.~L.~Liebling and C.~Palenzuela,
  Living Rev.\ Rel.\  {\bf 15}, 6 (2012)
  [arXiv:1202.5809 [gr-qc]].

\bibitem{Herdeiro:2016tmi} 
  C.~Herdeiro, E.~Radu and H.~Runarsson,
  Class.\ Quant.\ Grav.\  {\bf 33}, 154001 (2016)
  [arXiv:1603.02687 [gr-qc]].

\bibitem{Herdeiro:2015gia} 
  C.~Herdeiro and E.~Radu,
  Class.\ Quant.\ Grav.\  {\bf 32}, 144001 (2015)
  [arXiv:1501.04319 [gr-qc]].

\bibitem{Herdeiro:2014goa} 
  C.~A.~R.~Herdeiro and E.~Radu,
  Phys.\ Rev.\ Lett.\  {\bf 112}, 221101 (2014)
  [arXiv:1403.2757 [gr-qc]].

\bibitem{Herdeiro:2014ima}
  C.~A.~R.~Herdeiro and E.~Radu,
  Int.\ J.\ Mod.\ Phys.\ D {\bf 23},  1442014 (2014)
  [arXiv:1405.3696 [gr-qc]].
  
\bibitem{corona1} 
  G.~Matt, G.~C.~Perola and L.~Piro,
  Astron.\ Astrophys.\  {\bf 247}, 25 (1991).

\bibitem{corona2} 
  A.~Martocchia and G.~Matt,
  Mon.\ Not.\ Roy.\ Astron.\ Soc.\  {\bf 282}, L53 (1996).    
  
\bibitem{iron1} 
  L.~W.~Brenneman and C.~S.~Reynolds,
  Astrophys.\ J.\  {\bf 652}, 1028 (2006)
  [astro-ph/0608502].  
  
\bibitem{iron2} 
  C.~S.~Reynolds,
  Space Sci.\ Rev.\  {\bf 183}, 277 (2014)
  [arXiv:1302.3260 [astro-ph.HE]].   
  
\bibitem{iron3} 
  C.~Bambi, J.~Jiang and J.~F.~Steiner,
  Class.\ Quant.\ Grav.\  {\bf 33}, 064001 (2016)
  [arXiv:1511.07587 [gr-qc]].    
  
\bibitem{iron4}
  C.~Bambi, A.~Cardenas-Avendano, T.~Dauser, J.~A.~Garcia and S.~Nampalliwar,
  arXiv:1607.00596 [gr-qc].  
  
\bibitem{icb1} 
  J.~Jiang, C.~Bambi and J.~F.~Steiner,
  Astrophys.\ J.\  {\bf 811}, 130 (2015)
  [arXiv:1504.01970 [gr-qc]].  
  
\bibitem{icb2} 
  A.~Cardenas-Avendano, J.~Jiang and C.~Bambi,
  Phys.\ Lett.\ B {\bf 760}, 254 (2016)
  [arXiv:1603.04720 [gr-qc]].  
  
\bibitem{icb3} 
  Y.~Ni, J.~Jiang and C.~Bambi,
  JCAP {\bf 1609}, 014 (2016)
  [arXiv:1607.04893 [gr-qc]].    
  
\bibitem{dauser}
  T.~Dauser, J.~Garcia, J.~Wilms, M.~Bock, L.~W.~Brenneman, M.~Falanga, K.~Fukumura and C.~S.~Reynolds,
  Mon.\ Not.\ Roy.\ Astron.\ Soc.\  {\bf 430}, 1694 (2013)
  [arXiv:1301.4922 [astro-ph.HE]].    
  
\bibitem{c1} 
  C.~Bambi,
  Astrophys.\ J.\  {\bf 761}, 174 (2012)
  [arXiv:1210.5679 [gr-qc]].  
  
\bibitem{c2} 
  C.~Bambi,
  Phys.\ Rev.\ D {\bf 87}, 023007 (2013)
  [arXiv:1211.2513 [gr-qc]].    
  
\bibitem{p1} 
  M.~Zhou, A.~Cardenas-Avendano, C.~Bambi, B.~Kleihaus and J.~Kunz,
  Phys.\ Rev.\ D {\bf 94}, 024036 (2016)
  [arXiv:1603.07448 [gr-qc]].  
  
\bibitem{p4} 
  M.~Ghasemi-Nodehi and C.~Bambi,
  Phys.\ Rev.\ D {\bf 94}, 104062 (2016)
  [arXiv:1610.08791 [gr-qc]].    
  
\bibitem{relline} 
  T.~Dauser, J.~Wilms, C.~S.~Reynolds and L.~W.~Brenneman,
  Mon.\ Not.\ Roy.\ Astron.\ Soc.\  {\bf 409}, 1534 (2010)
  [arXiv:1007.4937 [astro-ph.HE]].    
  
\bibitem{Cunha:2015yba} 
  P.~V.~P.~Cunha, C.~A.~R.~Herdeiro, E.~Radu and H.~F.~Runarsson,
  Phys.\ Rev.\ Lett.\  {\bf 115}, 211102 (2015)
  [arXiv:1509.00021 [gr-qc]].

\bibitem{Vincent:2016sjq} 
  F.~H.~Vincent, E.~Gourgoulhon, C.~Herdeiro and E.~Radu,
  Phys.\ Rev.\ D {\bf 94}, 084045 (2016)
  [arXiv:1606.04246 [gr-qc]].

\bibitem{Franchini:2016yvq} 
  N.~Franchini, P.~Pani, A.~Maselli, L.~Gualtieri, C.~A.~R.~Herdeiro, E.~Radu and V.~Ferrari,
  arXiv:1612.00038 [astro-ph.HE].

\end{thebibliography}
\end{document}